# Observation of Nondegenerate Two-Photon Gain in GaAs

Matthew Reichert,[1,*] Arthur L. Smirl,[2] Greg Salamo,[3] David J. Hagan,[1,4] and Eric W. Van Stryland[1,4,†]

[1]*CREOL, College of Optics and Photonics, University of Central Florida, Orlando, Florida 32816, USA*
[2]*Laboratory for Photonics and Quantum Electronics, University of Iowa, Iowa City, Iowa 52242, USA*
[3]*Institute for Nanoscience & Engineering, University of Arkansas, Fayetteville, Arkansas 72701, USA*
[4]*Department of Physics, University of Central Florida, Orlando, Florida 32816, USA*

Two-photon lasers require materials with large two-photon gain (2PG) coefficients and low linear and nonlinear losses. Our previous demonstration of large enhancement of two-photon absorption in semiconductors for very different photon energies translates directly into enhancement of 2PG. We experimentally demonstrate nondegenerate 2PG in optically excited bulk GaAs via femtosecond pump-probe measurements. 2PG is isolated from other pump induced effects through the difference between measurements performed with parallel and perpendicular polarizations of pump and probe. An enhancement in the 2PG coefficient of nearly two orders-of-magnitude is reported. The results point a possible way toward two-photon semiconductor lasers.

Multi-photon processes, predicted by Dirac [1], have been understood theoretically since the foundational work of Göppert-Mayer [2]. The simultaneous absorption or emission of two quanta by a single electronic transition have come to be known as two-photon absorption (2PA) and two-photon emission (2PE), respectively. 2PE may be spontaneous or stimulated by either one or two photons, referred to as singly- and doubly-stimulated [3]. Utilizing doubly-stimulated 2PE, or two-photon gain (2PG), to realize a two-photon laser (2PL) has been a goal of nonlinear optics since shortly after the invention of the original laser [4,5]. 2PL's are predicted to have many desirable properties due to the inherent nonlinearity of the gain, including pulse compression [6], self-mode-locking [7], and unique photon statistics [8,9]. While observation of 2PG has been reported in a variety of materials [3,10-13], the development of 2PL's remains largely elusive. This is because 2PG is a weak process, and gain media suffer from various completing processes, e.g., excited state absorption, one-photon lasing, and parametric wave mixing [3]. Thus far, one group has demonstrated two-photon lasers using atomic barium [14] and potassium [15]. These, however, involved transitions from dressed states rather than eigenstates of unperturbed atoms [3]. All experimental work on 2PG thus far has focused only on the degenerate case where both emitted photons have the same frequency. However, there is no restriction on the energies of the individual photons, so long as their sum matches the energy difference between the initial and final states. Thus, 2PG may occur for both degenerate (D) and nondegenerate (ND) photon pairs.

Semiconductors have been proposed as potential 2PG media [16], and recent investigations have shown D-2PG in bulk AlGaAs waveguides [13]. It has been shown theoretically [17] and experimentally [18] that 2PA in bulk semiconductors may be increased by over two orders of magnitude when using photons of very different energies, so called extremely nondegenerate (END) photon pairs, e.g., $\hbar\omega_a \approx 10\hbar\omega_b$. This enhancement has been utilized for applications such as mid-IR detection [19] and 3D imaging [20]. Enhancement of 2PA translates into enhancement of 2PG, as the relation between the absorption and stimulated emission in the two-photon case is the same as for the single photon case. The two are inverse processes that depend on the populations of the initial and final states; with population inversion, 2PA changes directly into 2PG. Demonstration of an enhancement in 2PG may be a significant step in realizing a semiconductor 2PL. In this letter we experimentally demonstrate ND-2PG in optically excited bulk GaAs via femtosecond pump-probe experiments. This comprises, to the best of our knowledge, the first report of ND-2PG.

ND-2PG involves the interaction of two beams in the gain medium. In the pump-probe experiments presented here, the pump beam is much stronger than the probe, which is kept very weak to avoid self-induced nonlinearities. The evolution of the probe irradiance (within thin sample [21] and slowly-varying envelope [22] approximations) is governed by

$$\frac{\partial I_a}{\partial z} = 2\gamma_2(\omega_a;\omega_b)I_aI_b \quad (1)$$

where $I_a$ and $I_b$ are the irradiance distributions of the probe and pump, respectively, $\gamma_2$ is the 2PG coefficient, and the factor of 2 results from interference between the two beams [23]. The frequency arguments of $\gamma_2(\omega_a;\omega_b)$ indicate gain in the beam at $\omega_a$ due to the presence of the beam at $\omega_b$. The 2PG coefficient is related to the 2PA coefficient at equilibrium $\alpha_{2,0}$ by

$$\gamma_2(\omega_a;\omega_b) = \alpha_{2,0}(\omega_a;\omega_b)(f_c - f_v) \quad (2)$$

where, $f_c$ and $f_v$ are the Fermi-Dirac distributions describing the occupation of the conduction and valence bands,

respectively, evaluated where $E_c(\mathbf{k}) - E_v(\mathbf{k}) = \hbar\omega_a + \hbar\omega_b$. The 2PG coefficient varies from $-\alpha_{2,0}$ at equilibrium ($f_c = 0$, $f_v = 1$), to $\alpha_{2,0}$ with complete population inversion ($f_c = 1$, $f_v = 0$).

Experiments are conducted on samples consisting of a 4 μm layer of GaAs clad on both sides by ~1 μm of $Al_{0.47}Ga_{0.53}As$ to protect the surfaces and minimize surface recombination. $Al_{0.47}Ga_{0.53}As$ has a larger band-gap energy than GaAs (1.99 eV vs. 1.42 eV [24]) and is transparent at all wavelengths used. Samples are grown by molecular beam epitaxy on a semi-insulating GaAs substrate that is polished and etched away [25], and then glued (NOA81, Norland Products) to a sapphire substrate.

A Ti:sapphire chirped pulse amplifier system (Legend Elite Duo+ HE, Coherent) producing 12 mJ pulses at 800 nm of 35 fs duration (FWHM) at a 1 kHz repetition rate is used to generate all the beams involved in the experiment. A portion is split off and used directly as an excitation to optically generate carriers in the GaAs. The excitation energy at the sample is controlled via a half-wave plate and polarizer. Approximately 1.4 mJ pumps an optical parametric generator-amplifier (TOPAS-800, Light Conversion) and the difference between the signal and idler frequencies is generated in an external $AgGaS_2$ module. The difference frequency is used as the pump, and is tuned to 7.75 μm (160 meV) and spectrally filtered with a mid-IR bandpass filter (BPF-7730-180, Iridian Spectral Technologies) to narrow the bandwidth to 195 nm (4 meV, FWHM). A pair of wire-grid polarizers are used to both control the energy and polarization of the pump. Approximately 5 μJ of the 800 nm fundamental is focused into a 5 mm thick sapphire plate to generate a white-light continuum (WLC), which is spectrally filtered using narrow band filters (10 nm) and used as the probe. A calcite polarizer ensures linear polarization of the probe, which is rotated by a half-wave plate.

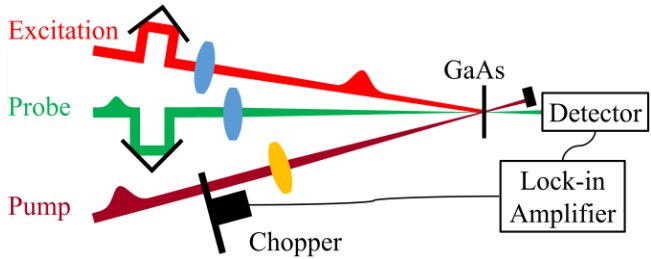

FIG. 1. Illustration of excite-pump-probe experiment.

The experimental setup is shown in Fig. 1. The 800 nm excitation pulse is sent to a delay line, where the timing is set such that it arrives at the sample 1.2 ps before the pump pulse. Optically excited electron-hole pairs are generated with an average energy 130 meV greater than the band-gap energy, and relax to the band edges within this time to produce population inversion (verified by one-photon gain experiments [26]). Spot sizes of the excitation, pump, and probe are 220 μm, 30 μm, and 20 μm ($HW1/e^2M$), respectively, as measured by knife-edge scans. The excitation spot size is much greater than those of both the pump and probe such that the carrier density is approximately uniform over their extent. The temporal delay between the pump and probe is controlled by a second computer-controlled delay line in the probe beam path.

In addition to 2PG, other physical mechanisms may cause changes in the probe transmission. The large carrier concentration needed to achieve population inversion induces significant free-carrier absorption (FCA), which is expected to dwarf the 2PG. We therefore employ lock-in techniques to distinguish between transmission changes from the pump and from the excitation. The probe transmission is detected via a lock-in amplifier (SR830, Stanford Research Systems) synchronous to the modulation frequency of a mechanical chopper in the pump beam. The measured signal is proportional to the change in the probe transmission that is induced only by the pump, thus eliminating the effect of the excitation alone.

A probe wavelength of $\lambda_a = 977$ nm ($\hbar\omega_a = 1.27$ eV) is initially selected, giving a transition energy $\hbar\omega_a + \hbar\omega_b = 1.43$ eV, and a nondegeneracy $\hbar\omega_a/\hbar\omega_b = 7.9$. For this combination of photon energies, the theoretically predicted nondegenerate enhancement of $\alpha_2$ is 71× the degenerate case for the energy sum [17]. Pump-probe measurements without excitation, i.e., without excited carriers, are shown in Fig. 2(a) for polarizations of the probe both parallel (black) and perpendicular (red) to the pump. The signal has been normalized by

$$T_N = 1 - \frac{S}{T}, \quad (3)$$

where $T_N$ is the normalized transmission, $S$ is the measured signal, and $T$ is the linear transmission in the absence of the pump. 2PA is a (nearly) instantaneous process occurring only while the two pulses are overlapped in time. Linearity of the transmission change with pump irradiance is verified by repeated measurements at several pump energies (not shown). Fits with Eqs. (1) and (2) give $\alpha_{2,\parallel} = (38 \pm 8)$ cm/GW and $\alpha_{2,\perp} = (16 \pm 4)$ cm/GW for parallel and perpendicular polarizations, respectively. These values, as well as the ratio $\alpha_{2,\parallel}/\alpha_{2,\perp} = 2.4 \pm 0.1$, are consistent both with theory [17,27] and previous measurements [18,28]. Although the nondegenerate enhancement increases the $\alpha_{2,0}$, the values measured here are not particularly large because the photon energy sum is near the band edge, where the density of states is low and degenerate $\alpha_{2,0}$ is small.

The excitation is then added with a peak fluence of 2.8 mJ/cm$^2$, and the pump-probe measurements are repeated, as shown in Fig. 2(b). The signal is again normalized via Eq. (3), but $T$ is now measured in the presence of the excitation. The probe transmission is now increased at zero delay, consistent with 2PG. There is also, however, a reduction in transmission at positive delay, where the pump arrives before the probe and they no longer overlap within the sample. The pump induces a change in absorption on a much longer time scale. This may be due to carrier heating resulting in additional FCA of the probe [29]. The polarization dependence of the measured signal varies with delay; the difference between parallel and perpendicular polarizations is greatest at zero delay and disappears at positive delay. This indicates that the mechanism responsible for the reduced transmission at positive delay is different from that for the increase in transmission at zero delay.

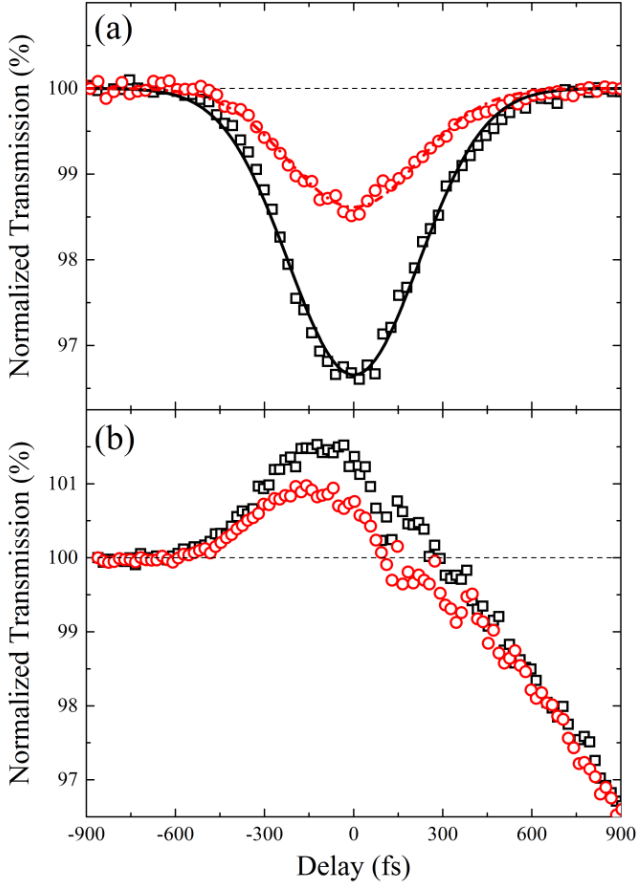

FIG. 2. Pump-probe measurements (a) without excitation showing 2PA for both (black squares) parallel and (red circles) perpendicular polarizations of pump and probe and (b) with an excitation fluence of 2.8 mJ/cm$^2$ showing both 2PG and pump induced FCA. Curves in (a) show fits to extract $\alpha_{2,0}$.

The total response may be described as the sum of two mechanisms having different polarization dependences. 2PG has the same polarization dependence as 2PA, while the second component is isotropic, i.e., independent of polarization direction, which is consistent with FCA. Furthermore, 2PG can only occur when the two pulses are overlapped within the sample, while the second mechanism occurs over a much a longer timescale. The propagation of the probe in the presence of ND-2PG and isotropic absorption is

$$\frac{\partial I_a}{\partial z} = 2\gamma_{2,\parallel} I_a I_b - \alpha_{\text{iso}} I_a \quad (4)$$

for parallel polarizations, and

$$\frac{\partial I_a}{\partial z} = 2\gamma_{2,\perp} I_a I_b - \alpha_{\text{iso}} I_a \quad (5)$$

for perpendicular polarizations, where $\alpha_{\text{iso}}$ is the isotropic absorption coefficient. Taking the difference between Eqs. (4) and (5) eliminates the isotropic absorption term, while retaining the 2PG,

$$\left.\frac{\partial I_a}{\partial z}\right|_\parallel - \left.\frac{\partial I_a}{\partial z}\right|_\perp = 2(\gamma_{2,\parallel} - \gamma_{2,\perp}) I_a I_b. \quad (6)$$

Hence, subtracting the measurements with parallel and perpendicular polarizations, which we define as $\Delta T$, eliminates the induced change in isotropic absorption. The result both with and without the excitation generating a population inversion is shown in Fig. 3. $\Delta T$ is only nonzero about zero delay, and follows the same cross-correlation shape in both cases. This validates the assumption that a mechanism of isotropic symmetry, such as FCA, is responsible for the reduction in transmission at positive delay. The increase in transmission at zero delay is therefore due to ND-2PG. The previously measured ratio $\alpha_{2,\parallel}/\alpha_{2,\perp}$ applies to the case of 2PG as well, and gives $\gamma_{2,\perp} = \gamma_{2,\parallel}/2.4$, allowing the use of Eq. (5) to determine the 2PG coefficient. The solid blue curve in Fig. 3 corresponds to a fit with Eq. (6) yielding $\gamma_{2,\parallel} = (14 \pm 5)$ cm/GW.

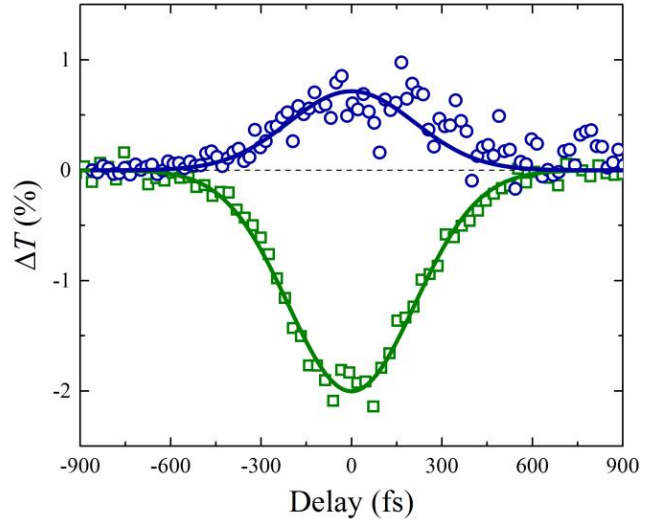

FIG. 3. Difference between parallel and perpendicular polarizations both without (green squares) and with (blue circles) excitation. Curves represent fits for $\gamma_{2,\parallel}$.

Conversion from 2PA to 2PG requires creation of a population inversion in the specific region of k-space where the transition occurs (see Eq. (2)). Measurements of $\Delta T$ at zero pump-probe delay is plotted vs. excitation fluence in Fig. 4, showing the transition from absorption to gain with population inversion. Without the excitation, $\Delta T = -2$ %, corresponding to 2PA. As the excitation fluence is increased, the carrier population near the band edge grows, causing a reduction in the 2PA, which eventually changes into 2PG above 1 mJ/cm$^2$. At still higher fluences, growth of the 2PG is limited by absorption saturation of the excitation, reaching a maximum of $\Delta T = 0.7$ %. The maximum observed value of value of $\gamma_2 < \alpha_{2,0}$, indicating incomplete population inversion. This is due to the finite temperature of the excited electrons and holes, which are estimated to have a carrier temperature ~600 K [26]. Even with this incomplete population inversion, the measured value of $\gamma_{2,\parallel} = (14 \pm 5)$ cm/GW is still ~26× $\alpha_{2,0}$ in the degenerate case at the same photon energy sum [17]. Enhancement over D-2PG, however, is independent of the population inversion, and exhibits the same 71× increase as ND-2PA.

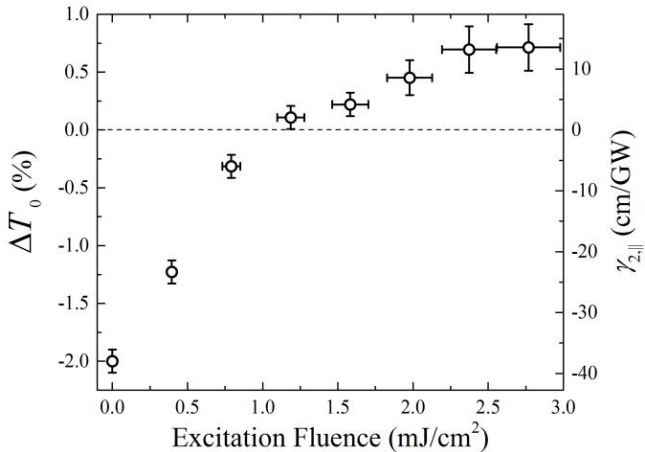

FIG. 4. Difference in pump induced transmission change between parallel and perpendicular polarization at zero delay versus excitation fluence. Right axis shows 2PG coefficient determined from Eq. (5).

The spectrum of 2PG depends on the energy distribution of carriers within the bands; inversion decreases higher in the band, resulting in a reduced $\gamma_2$. To demonstrate this dependence, pump-probe measurements are repeated at larger photon energy sums (transition energies), with probe wavelengths of 964 nm and 947 nm, and $\gamma_{2,\parallel}$ is determined. To isolate the effects of carrier distribution from the density of state dependence, the ratio $\gamma_{2,\parallel}/\alpha_{2,\parallel} = f_c - f_v$ is plotted in Fig. 5, which depends only of the population inversion. The result shows only inversion near the band edge, with the expected reduction in inversion with increasing energy.

In conclusion, we have experimentally demonstrated ND-2PG in bulk GaAs via pump-probe methods. Measurements relied on the different tensor symmetries of 2PG and the observed long lived reduction in absorption that allows for their separation with measurements at different polarizations. This constitutes, to the best of our knowledge, the first observation of ND-2PG in any medium. The intermediate state resonance enhancement responsible for increasing 2PA for extremely different photon energies translates directly to enhanced 2PG. Utilization of this enhancement may open the possibility of END two-photon semiconductor lasers (2PSL's). We note, however, that we have not demonstrated net two-photon gain that exceed the various losses. Therefore, to determine if such a 2PSL is possible, losses from competing processes including FCA, and three-photon absorption must be considered in detail.

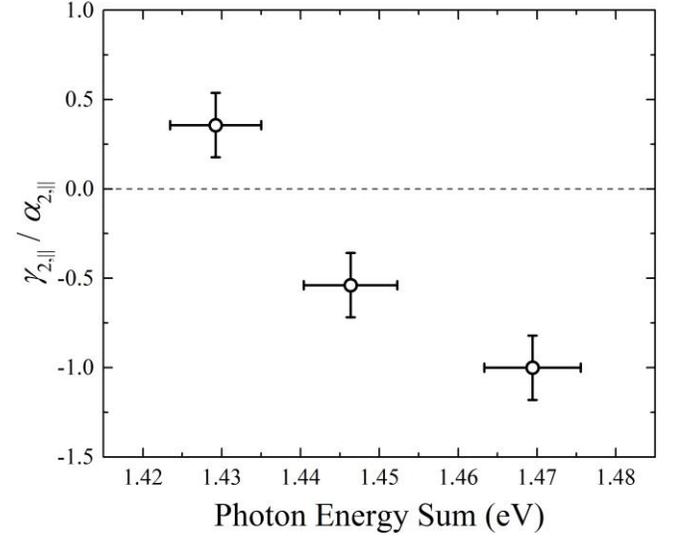

FIG. 5. Plot of $\gamma_{2,\parallel}$ measured with an excitation fluence of 2.8 mJ/cm$^2$ divided by $\alpha_{2,\parallel}$ without excitation versus photon energy sum (transition energy).

## ACKNOWLEDGMENTS

We thank Vasyl P. Kunets at the University of Arkansas for fabricating the GaAs samples. This work was supported by the National Science Foundation grants ECCS-1202471 and ECCS- 1229563.

* Current address: Department of Electrical Engineering, Princeton University, Princeton, New Jersey 08455, USA, Electronic address: mr22@princeton.edu
† Electronic address: ewvs@creol.ucf.edu